\documentclass[aps,twocolumn,superscriptaddress]{revtex4}
\usepackage{graphicx}
\usepackage{times}
\usepackage{amssymb,amsmath}

\begin{document}
\title{Experimental generation of single photons via active multiplexing}

\author{Xiao-song Ma}
\affiliation{Institute for Quantum Optics and Quantum Information (IQOQI), Austrian
Academy of Sciences, Boltzmanngasse 3, A-1090 Vienna, Austria}
\affiliation{Faculty of Physics, University of Vienna, Boltzmanngasse 5, A-1090 Vienna,
Austria}\email{xiaosong.ma@univie.ac.at}
\author{Stefan Zotter}
\affiliation{Institute for Quantum Optics and Quantum Information (IQOQI), Austrian
Academy of Sciences, Boltzmanngasse 3, A-1090 Vienna, Austria}
\affiliation{Faculty of Physics, University of Vienna, Boltzmanngasse 5, A-1090 Vienna,
Austria}
\author{Johannes Kofler}
\affiliation{Institute for Quantum Optics and Quantum Information (IQOQI), Austrian
Academy of Sciences, Boltzmanngasse 3, A-1090 Vienna, Austria}
\affiliation{Faculty of Physics, University of Vienna, Boltzmanngasse 5, A-1090 Vienna,
Austria}
\author{Thomas Jennewein}
\affiliation{Institute for Quantum Optics and Quantum Information (IQOQI), Austrian
Academy of Sciences, Boltzmanngasse 3, A-1090 Vienna, Austria}
\affiliation{Institute for Quantum Computing and Department of Physics and Astronomy, University of Waterloo, 200 University Avenue West, Waterloo, Ontario, N2L 3G1, Canada}
\author{Anton Zeilinger}
\affiliation{Institute for Quantum Optics and Quantum Information (IQOQI), Austrian
Academy of Sciences, Boltzmanngasse 3, A-1090 Vienna, Austria}
\affiliation{Faculty of Physics, University of Vienna, Boltzmanngasse 5, A-1090 Vienna, Austria}
\affiliation{Vienna Centre for Quantum Science and Technology, Boltzmanngasse 3, A-1090, Vienna, Austria}
\email{anton.zeilinger@univie.ac.at}

\begin{abstract}
An on-demand single-photon source is a fundamental building block in quantum science and technology. We experimentally demonstrate the proof of concept for a scheme to generate on-demand single photons via actively multiplexing several heralded photons probabilistically produced from pulsed spontaneous parametric down-conversions (SPDCs). By utilizing a four-photon-pair source, an active feed-forward technique, and an ultrafast single-photon router, we show a fourfold enhancement of the output photon rate. Simultaneously, we maintain the quality of the output single-photon states, confirmed by correlation measurements. We also experimentally verify, via Hong-Ou-Mandel interference, that the router does not affect the indistinguishability of the single photons. Furthermore, we give numerical simulations, which indicate that photons based on multiplexing of four SPDC sources can outperform the heralding based on highly advanced photon-number-resolving detectors. Our results show a route for on-demand single-photon generation and the practical realization of scalable linear optical quantum information processing.
\end{abstract}

\maketitle

Numerous applications for on-demand single-photon sources have been proposed in the field of quantum-enhanced science and technology~\cite{Gisin2002,Scarani2009,Bouwmeester1997, Pan1998, Jennewein2002, Kok2007}. The commonly studied methods to generate on-demand single photons are based on emissions from molecules~\cite{Lounis2000, Lettow2010}, atoms~\cite{Chou2004}, color centers in diamond~\cite{Kurtsiefer2000}, quantum dots~\cite{Michler2000, Santori2002}, and donor impurities~\cite{Sanaka2009}. However, each of these methods has certain challenges to overcome. For atomic systems, the repetition rates, the collection efficiency, and the complexity of the experimental setup are the main obstacles. For color centers, quantum dots and donor impurities, it is difficult to achieve a time-bandwidth limited emission spectrum~\cite{Batalov2008} and indistinguishable output photons from different sources~\cite{Patel2010}.

An alternative approach is the
so-called heralded single-photon source (HSPS) based on the process of spontaneous
parametric down-conversion (SPDC), where a pair of photons is
created from a nonlinear crystal pumped by a laser. The detection of one trigger photon
indicates, or heralds, the presence of its twin. Because of the
phase-matching condition and energy conservation, a HSPS from SPDC has some appealing features. The linear momentum (direction), polarization, and wavelength of
that heralded photon are well defined via the measurement of its twin. The
scheme faces two main challenges: (1) The conversion process is inherently random
and hence there is no prior knowledge of when the heralding event
will occur. (2) Because of the nature of the SPDC process, in
addition to the probability of generating one pair of photons, $P_{1}$, there is also a finite probability of generating more
than one pair, $P_{>1}$, via higher order emissions, which decreases the quality of the HSPS. This latter generation
probability increases nonlinearly with the interaction strength of the pump laser and the nonlinear crystal.

To surmount the random production of photon pairs from SPDC, a pulsed laser can be used to pump the nonlinear crystals, and hence the photon pairs can be created only at certain times. When the pulse duration of the pump laser is much shorter than the coincidence measurement time, the probability of generating \textit{n} pairs of photons per pulse can be represented as a Poisson distribution:
\begin{equation}
P_{n}=\frac{\bar{N}^{n}e^{-\bar{N}}}{n!},  \textrm{ for } n=0,1,2,...,\label{Poisson}
\end{equation}%
where the mean number of pairs, $\bar{N}$, depends on the pump power and parameters of the crystals. Note that it has been shown that to use Bose (thermal) distribution, the analysis presented below would yield essentially the same results~\cite{Migdall2002}. An ideal HSPS would have exactly one photon generated from each pulse, which requires $\bar{N}=1$. However, for $\bar{N}=1$, the probability to observe more than one photon is 0.26 due to the Poisson distribution. This significant higher order emissions probability, $P_{>1}$, drastically reduces the quality of the HSPS. For this reason, pulsed SPDC is normally operated at low power (thus $P_{>1}\ll1$) and obeys a trade-off between the count rate and quality of the HSPS. This is the reason why a single SPDC source is fundamentally limited. This limitation can be overcome by using spatial or temporal multiplexing of several sources~\cite{Migdall2002, Shapiro2007, Pittman2002, Jeffrey2004, McCusker2009}. Although this idea has attracted significant attention, as highlighted in recent review articles~\cite{Ladd2010, Kok2010}, it has not yet been experimentally realized. Here we follow and extend the proposal in Ref.\ [16]. We report an experimental realization of a 4-SPDC single-photon multiplexing system and demonstrate a proof-of-principle enhancement over a 1-SPDC single-photon source.

\section*{Theoretical analysis}
\begin{figure}
\includegraphics[width=0.49\textwidth]{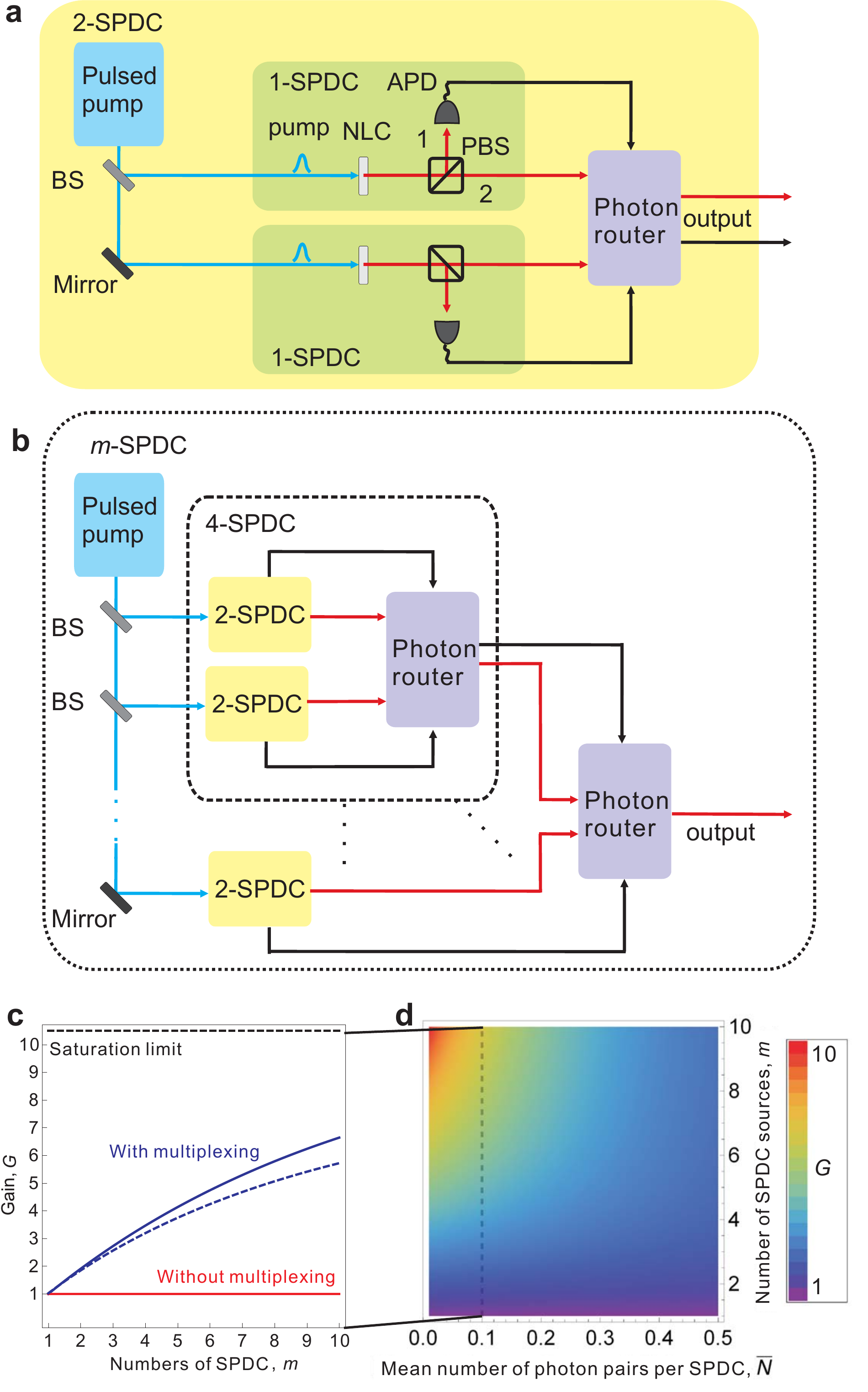}
\caption{\label{fig1}(Color online) Scheme and calculations of generating on-demand single photons via multiplexing with active feed forward from spontaneous parametric down conversion (SPDC). (a) A 2-SPDC module consists of two 1-SPDC sources coupled with a photon router. Each 1-SPDC source produces a pair of photons probabilistically from a nonlinear crystal (NLC) pumped by laser pulses whose power is split by a beam splitter (BS). The signal of each avalanche photodiode (APD) detector, indicated by the black arrows, is feed-forwarded and controls the photon router to direct one of the successfully created photons to a single output. This output single photon along with the output feed-forward signal can then be used in the next level. Suitable individual optical delays (not shown) are incorporated in order to erase the output photons' temporal distinguishability from different sources. (b) Similarly, a 4-SPDC multiplexing system can be realized by coupling two 2-SPDC modules with another photon router. An $m$-SPDC multiplexing system requires $m$ SPDC sources and $(m-1)$ routers. (c) The calculated gain, \textit{G}, of using a multiplexed system is plotted as a function of the number of SPDC sources, $m$, where only integer values of $m$ is relevant (blue solid line). The trivial gain of using a 1-SPDC source is shown for comparison (red solid line). The mean photon pair number per SPDC source is $\bar{N}=0.1$, indicated with the vertical dashed line in (d), and the horizontal black dashed line represents the saturation limit $\textrm{lim}_{m\rightarrow\infty}G=\frac{1}{1-e^{-\bar{N}}}\approx10.5$. In addition, we show the gain when using multiplexing with a 0.95 transmission per path router and 0.97 of the polarization router (blue dashed line) in order to illustrate loss effects. For details, see the main text and Appendix A. \textbf{d}, Calculation result of the gain, \textit{G} as a function of both the numbers of SPDC sources $m$ and the mean photon pair number per SPDC source $\bar{N}$.}
\end{figure}

In the spatial multiplexing scheme, $m$ SPDC sources are pumped by a single pulsed laser whose power is split equally for each SPDC source by a series of beam splitters (BSs). These SPDC sources are then coupled by fast photon routers and directed to a single output. Each individual SPDC source has the probability to generate one pair, $P_{1}$. The pump power input is chosen low enough that the generation of more than one pair, $P_{>1}$, is much smaller than $P_{1}$. With sufficiently large $m$ and feed-forward operation of the fast photon router, the probability of obtaining one pair emission in this array, $Q_{1}$, can approach unity and hence be on demand. The detection signal of the heralding (trigger) photon of an individual source is used to identify which source has produced a pair of photons and to control the photon routers to direct the successfully created photon to the single output of this $m$-SPDC array. Shapiro \textit{et al.}\ proposed a
modular configuration by using $2\times1$ routers~\cite{Shapiro2007}, where two inputs can be routed to a single output. As shown in Fig.\ \ref{fig1}(a), in a 1-SPDC HSPS, one nonlinear crystal (NLC) cut for collinear type-II
phase matching is pumped by laser pulses, where a pair of
orthogonally polarized photons is generated. Their quantum state
is a product state in the polarization degree of freedom: $|\Phi\rangle_{\textrm{12}}= \left\vert
V\right\rangle _{\textrm{1}}\left\vert H\right\rangle_{\textrm{2}}$. $\left\vert V\right\rangle_{\textrm{1}}$ and $\left\vert H\right\rangle_{\textrm{2}}$ denote the vertical and horizontal polarization states of photons 1 and 2 respectively.
Then they are separated by a polarizing beam splitter (PBS). The
detection of the vertically polarized photon heralds the existence of the horizontally polarized photon, which is the output of this 1-SPDC source. Two
such 1-SPDC sources can be coupled with a photon router and integrated into
a 2-SPDC module, as also shown in Fig.\ \ref{fig1}(a). It is straightforward to increase the size of the network. Two 2-SPDC modules can build a 4-SPDC module and so on [Fig.\ \ref{fig1}(b)]. The advantages of this multiplexing are as follows: First, one can enhance the one-pair-generation probability by a gain factor of (see Appendix A)
\begin{equation}
G=\frac{Q_{1}}{P_{1}}=\frac{1-P^{m}_{0}}{1-P_{0}}.\label{Gain}
\end{equation}%
Second, the signal-to-noise ratio of an $m$-SPDC array is the same as that of a 1-SPDC source: $\frac{Q_{1}}{Q_{>1}}=\frac{P_{1}}{P_{>1}}$. $Q_{>1}$ is the probability of obtaining more than one pair in this array. In Fig.\ \ref{fig1}(c), the gain $G$ of a multiplexed system with $m$ sources (blue) is shown and compared with a single 1-SPDC source (red) at $\bar{N}=0.1$. For $P_0$ close to 1 ($\bar{N}$ close to 0), it first increases linearly with $m$. The gain saturates at the value $\textrm{lim}_{m\rightarrow\infty}G=\frac{1}{1-P_{0}}=\frac{1}{1-e^{-\bar{N}}}$ for an infinite number of sources. The black dashed line in Fig.\ \ref{fig1}(c) shows this saturation limit at $\bar{N}=0.1$. The dependence of the gain on the mean photon pair number per SPDC source $\bar{N}$ and the numbers of the sources $m$ is shown in Fig.\ \ref{fig1}(d). Intuitively, in order to obtain high-quality and high-rate single photons, one should operate in a regime, in which $\bar{N}$ is low for each SPDC, in order to reduce the higher order emission per source, and $m$ is large, so that the chance of one source firing in this array is high. This is the reason why the optimal operating combination of $\bar{N}$ and $m$ in terms of maximizing $G$ lies in the upper-left corner in Fig.\ \ref{fig1}(d), given that $m$ is limited to 10.

\section*{Experiment}
Initially, we created photon pairs from a $\beta$-barium-borate crystal (BBO) via noncollinear type-II phase
matching~\cite{Kwiat1995}. As shown in Fig.\ 2a, photons 1 and 2 were
generated from BBO1 and their quantum state was
$|\Phi ^{-}\rangle_{\textrm{12}}=\frac{1}{\sqrt{2}}%
(\left\vert H\right\rangle _{\textrm{1}}\left\vert V\right\rangle _{\textrm{2}%
}-\left\vert V\right\rangle _{\textrm{1}}\left\vert H\right\rangle _{\textrm{2}%
}) $. In comparison with the product state generated from the
collinear phase matching ($\left\vert
V\right\rangle _{\textrm{1}}\left\vert H\right\rangle_{\textrm{2}}$), this non-collinear scheme
is equivalent to having two 1-SPDC sources ($\left\vert H\right\rangle
_{\textrm{1}}\left\vert V\right\rangle _{\textrm{2}}$ and $\left\vert
V\right\rangle _{\textrm{1}}\left\vert H\right\rangle _{\textrm{2}}$),
each subject to half of the pump power. Additionally, the non-collinear
type-II phase matching overlaps the spatial modes of these two
sources as schematically shown by the dashed PBS in Fig.\ 2(c). By using a polarization switch and a polarizer, these two 1-SPDC sources form a 2-SPDC module. In comparison with the scheme proposed in Ref.\ [16], our noncollinear scheme has two advantages: The amounts of crystals and beam splitters needed for building an $m$-SPDC array are reduced from $m$ to $\frac{m}{2}$ and there is no need to spatially overlap the photons generated from two 1-SPDC sources. An identical 2-SPDC module subsequently generated photons 3 and 4 from BBO2. See Appendix B for details.

\begin{figure}
\includegraphics[width=0.48\textwidth]{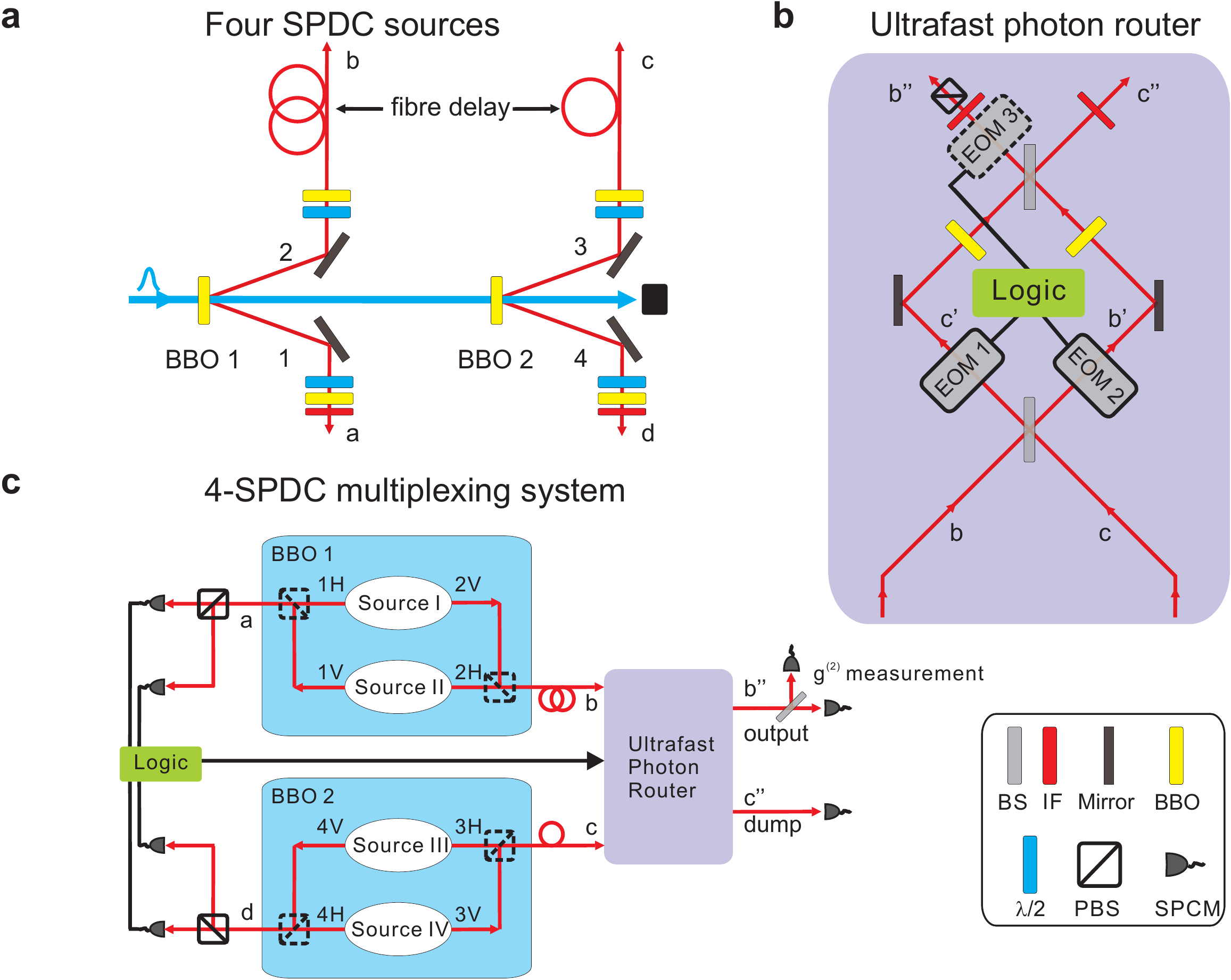}
\caption{\label{fig2}(Color online) Experimental setup. (a) Two pairs of the polarization entangled photons (1 and 2, 3 and 4) were generated from two $\beta$-barium borate crystals (BBO1 and BBO2) via SPDC, where half-wave plates ($\lambda$/2) and compensating BBO crystals were used to counter walkoff effects in the down-conversion crystal, and single-mode fibers and interference filters (IFs) were used to clean their spatial and spectral modes, respectively. Photons 1 and 4 entered into spatial modes a and d respectively. Photons 2 and 3 entered into spatial modes b and c and were delayed in single-mode fibers ($\approx 97 \textrm{m}$) with respect to photons 1
and 4. This allowed the logic and electronics to have enough time to
control the EOMs in the router. Due to the sequential configuration of the sources (see Appendix A), appropriate individual fiber delays are implemented to eliminate the output photons' temporal distinguishability from different sources. The layout of the ultrafast photon router is shown
in (b). See text and Appendix Cfor details and notations. (c) The
experimental scheme of a 4-SPDC multiplexed system with active
feed-forward. Sources I and II, III and IV generated photon pairs from BBO 1 (BBO 2). Sources I and II (III and IV) were inherently overlapped via the noncollinear phase matching, as explained in the text. The polarization states of photons 1 and 4 were measured with polarizing beam splitters (PBS) and single-photon counting modules (SPCMs).
The detection signals of photons 1
and 4 were fed into the logic and then used to control the ultrafast
photon router. Depending on whether photon 1 or photon 4 was
detected, the ultrafast photon router would be switched on or off. Thus, all the photons exited from mode b$''$ and no photons exited from c$''$.
The quality of the output single photons was quantified via measurements of the correlation function, $g^{2}(0)$.}
\end{figure}

These two 2-SPDC modules form a 4-SPDC multiplexed system with an ultrafast photon router, which satisfies two main criteria: (1) short response time, which is necessary for
the feed-forward operations in a large system, and (2) high routing visibility, which is needed for a high quality HSPS. As shown in Fig.\ 2(b), we employed a Mach-Zehnder interferometer (MZI) to realize the ultrafast photon router. It consisted of two
50:50 BSs, mirrors, and most importantly two electro-optic
modulators (EOMs) with one in each arm of the MZI, where two rubidium
titanate phosphate (RTP) crystals were used as the electro-optic
material. This design of the router has several advantages: (1) It can route incoming photons independent of their polarizations, which is crucial for certain tasks. In the schemes of polarization-entanglement-based quantum cryptography~\cite{Treiber2009} and quantum computation with polarization encoded photonic qubits~\cite{Kok2007}, any polarization-dependent routing will destroy the polarization entanglement and hence defeat the original goal of achieving communication security and computation speed-up provided by the entanglement. For instance, in polarization-entanglement-based quantum cryptography described in Ref.~\cite{Treiber2009}, the source is located at Alice's laboratory, and out of each pair of entangled photons, one is sent to Alice and the other to Bob. In a multiplexed scenario, many sources produce polarization-entangled photon pairs. The trigger photons are sent to and measured by Alice. Depending on the outcome of her polarization measurement, she sends an electronic signal to the routers for multiplexing the photons which will be sent to Bob. Multiplexed systems with our router can enhance the key rate while keeping the quantum bit error rate (due to the multiple pair generations) constant. Whatever critical key rate is needed at the receiver, $m$-SPDC multiplexing extends the possible communication distance between the two parties by the additional distance $\ln(m)L_0$ compared to 1-SPDC, where $L_0$ is the $1/e$ decay distance in the transmission channel (valid for both free space and fiber). In a 1550 nm fiber-based cryptography system, a 4-SPDC multiplexing system could increase the available distance by $\ln(4)\!\cdot\!21.7 \, \textrm{km}\approx 30.1 \, \textrm{km}$. Note however, whether one needs entangled photons depends on what kind of quantum-information processing task the photon will be used for later on. One does not need entanglement for the operation of the multiplexing procedure itself or for using multiplexing in the Bennett and Brassard quantum cryptography protocol~\cite{Bennett1984}, as the latter is not based on entanglement. Then one can also use a mixture of polarization-correlated product states as the input of the multiplexing scheme. (2) A similar design has been widely used in the photonic industry to implement intensity modulators~\cite{Reed2010}. Therefore, the scheme demonstrated here has the potential to be implemented with the mature technology of integrated optics.

The optical axes of both RTP crystals in the EOMs were
oriented along $45^{\circ}$ and the voltages applied to them were
always of the same amplitude but with opposite polarities. If the opposite half-wave voltages (HWVs) were applied to the EOMs, a $\pi$ phase shift would be
introduced to the MZI. We locked the phase of the MZI at $0$ when the
EOMs were off. This 0 phase is defined via the condition that all the photons that
enter from input b exit into c$''$, and c into b$''$ (Fig.\ 2b). Therefore, when the EOMs
were on and were applied with opposite HWVs, the photons that entered from input b exited into b$''$, and c into c$''$.

The rising and falling times of our router for a $\pi$-phase modulation are each about $5.6$ ns. (This could be reduced to tens of picoseconds with state-of-art integrated photonics devices~\cite{Reed2010}.) The recharger time of the high-voltage driver of our EOMs is about $50$ ns. The traveling time in the used coaxial cables is also a few nanoseconds. Together, this in principle allows a repetition rate of our router of approximately $15$ MHz. (Note that the dead time of trigger APDs is about $150$ ns, which will limit the operating frequency of small-scale multiplexing systems.) The visibility of $\pi$-phase modulation (routing visibility), $V^{\pi}$, was above 95\%. It is defined as $V^{\pi}_{\textrm{b}}=\frac{I^{\pi}_{\textrm{b}''}-I^{0}_{\textrm{b}''}}{I^{\pi}_{\textrm{b}''}+I^{0}_{\textrm{b}''}}$ for spatial mode b as the input, where $I^{\pi}_{\textrm{b}''}$ and $I^{0}_{\textrm{b}''}$ are the output intensities of spatial mode b$''$ of phase $0$ and $\pi$, respectively. Similar results were obtained for spatial mode c as the input. Therefore, this photon router meets the two requirements: short response time (compared to the recharging time of our EOMs) and high routing visibility. See Appendix C for further details of this ultrafast photon router.

The 4-SPDC multiplexing system [Fig.\ 2(c)] was operated as follows. We detected photons 1 and 4 in the modes of 1H, 1V, 4H and 4V, and sent the electrical signals from the detectors through a custom-built programmable logic to the router. When the detector on mode 1H or 1V fired, the EOMs in the MZI were switched on with HWV and thus the phase of the MZI was switched to $\pi$. Therefore, photon 2 was routed to spatial mode b$''$. When the detector on mode 4H or 4V fired, the EOMs in the MZI were switched off, and the phase of the MZI remained at 0. Therefore, photon 3 was routed to spatial mode b$''$. In the last step, an additional electro-optical modulator (EOM3) placed in spatial mode b$''$ acted as a polarization router and performed an active polarization rotation on the output photon, before it was finally cleaned by using a PBS. However, due to limited resources, we bypassed EOM3 and performed this operation with post-selection. Note that introducing EOM3 would only reduce the routing visibility by at most 1\% (with a polarization switching visibility of at least 99\%)~\cite{Scheidl2008}.


In order that our \textit{m}-SPDC source is useful in photonic quantum-information processing, it is most important that it performs well in Hong-Ou-Mandel (HOM) two-photon interference~\cite{Hong1987}, which is at the heart of many protocols and has been demonstrated with different systems and schemes. For instance, interference between consecutive photons emitted from the same quantum dots~\cite{Santori2002}, between photons from different donor impurities~\cite{Sanaka2009} and molecules~\cite{Lettow2010}, as well as between two independent SPDC sources~\cite{Kaltenbaek2006, Halder2007} have been shown. At present, in terms of the count rate and quality of the independent single photons, femtosecond-pulsed SPDC is still better than other schemes and has been successfully used in multiphoton experiments in the past. A single-photon source based on our scheme is suitable for high quality HOM two-photon interference and enables multiphoton experiments~\cite{Pan2008}. The present multiplexing scheme includes the photon router as the only additional complexity which was not involved in earlier experiments. Since the path length difference of the MZI can be controlled very accurately, the router will not introduce any temporal distinguishability, even though it is based on first-order interference which is wavelength sensitive (nm range). In contrast, most of the photonic quantum-computation experiments (C-phase gate~\cite{Langford2005, Kiesel2005, Okamoto2005}, entanglement swapping~\cite{Pan1998, Jennewein2002}, etc.) rely on second-order interference, which is coherence length sensitive (depends on the context, but at least more than $10^4$~nm). In order to empirically exclude possible degradations of the indistinguishability due to the router, we have performed HOM two-photon interference experiments with the router.

\begin{figure}
\includegraphics[width=0.49\textwidth]{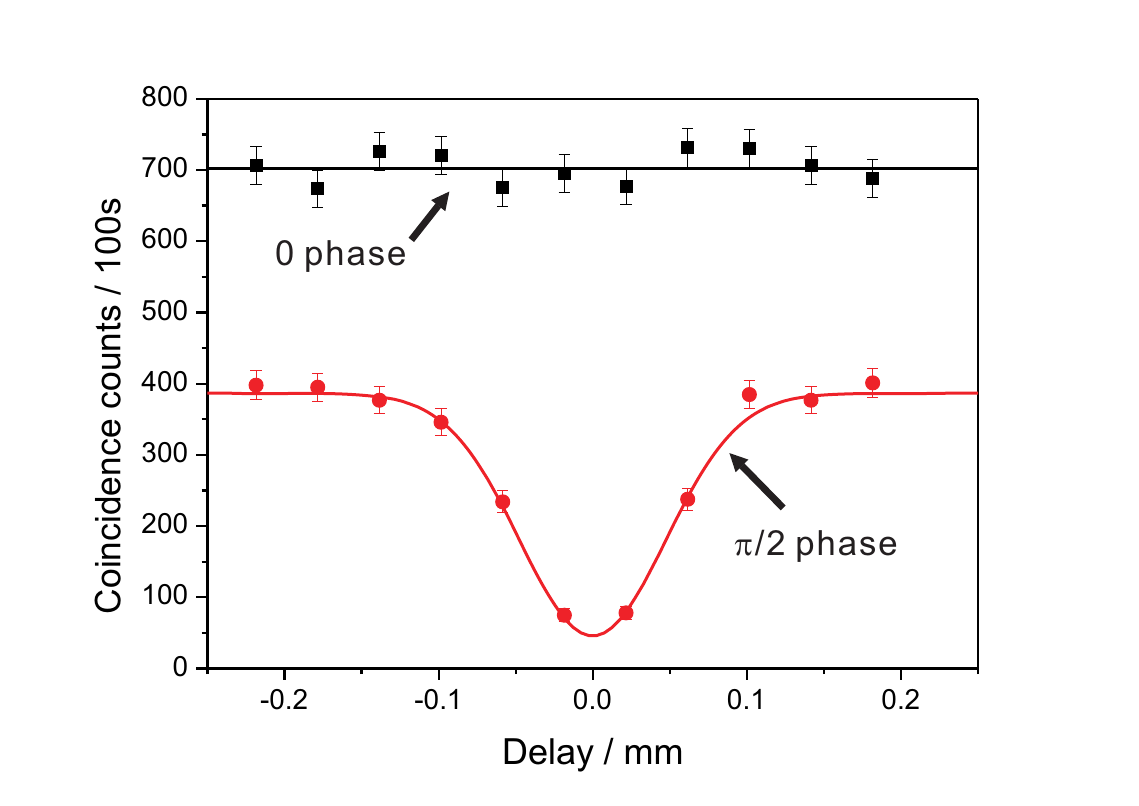}
\caption{\label{figHOM}(Color online) Experimental HOM two-photon interference performed with the photon router.
The two-fold coincidences are measured between two detectors placed directly behind two outputs of the photon router (b$''$ and c$''$ in Fig.\ \ref{fig2}b) and are plotted over the
relative delay of the interfering photons [realized by changing the length of path c in Fig.\ 2(b)]. By adjusting the voltage applied on the EOMs, we set the phase of the MZI and hence varied the reflectivity (and transmissivity) of the router. When the phase of the MZI is $\pi/2$, HOM two-photon interference shows up (red circles). This is because the distinguishability of the two input photons' spatial modes has been erased by the router. Consequently, the minimum coincidence counts occur for the temporal overlap of the two photons, i.e.\ zero time delay. The red solid line is a Gaussian fit to the data with a visibility of $88.7\%\pm3.8\%$. When the phase is 0, path length difference insensitive coincidence counts are obtained (black squares) and no HOM two-photon interference shows up, because the two photons remain distinguishable in their spatial modes. The black solid line is the average of the coincidence counts for 0 phase setting. Background has not been subtracted. Error bars indicate $\pm1$ standard deviation.}
\end{figure}

Experimentally, we use a pair of photons generated from one source and send it to the router. We vary the path-length difference between these two photons with a motorized translation stage mounted on one of the fiber coupling stages and measure the twofold coincidences between two detectors placed directly behind two outputs of the router [b$''$ and c$''$ in Fig.\ \ref{fig2}(b)]. The phase of the MZI and hence the reflectivity (and transmissivity) of the router is varied by applying different voltages on the EOMs. In case of a $\pi/2$ phase, the router becomes a balanced BS and the distinguishability of two input photons' spatial modes is erased. In consequence, the minimum of the coincidence counts occurs for the optimal temporal overlap (with the help of suitable individual fiber delays) of the two photons, and HOM two-photon interference with a visibility of $88.7\%\pm3.8\%$ can be observed (red circles in Fig.\ \ref{figHOM}). In case of a $0$ phase, the whole router represents a highly transmissive BS (routing visibility above 95\%) and the two photons remain distinguishable in their spatial modes. Correspondingly, one obtains path-length-difference insensitive coincidence counts (black square in Fig.\ \ref{figHOM}). This is in agreement with complementarity, where in principle no HOM interference can be observed. In addition, we have compared the results when using the router with those when using a normal free-space BS: quantitatively identical results have been obtained (within the error bars). These HOM experiments together with the earlier works~\cite{Kaltenbaek2006, Pan2008} demonstrate the suitability of the multiplexed photons for scalable quantum information processing.

\begin{figure}
\includegraphics[width=0.46\textwidth]{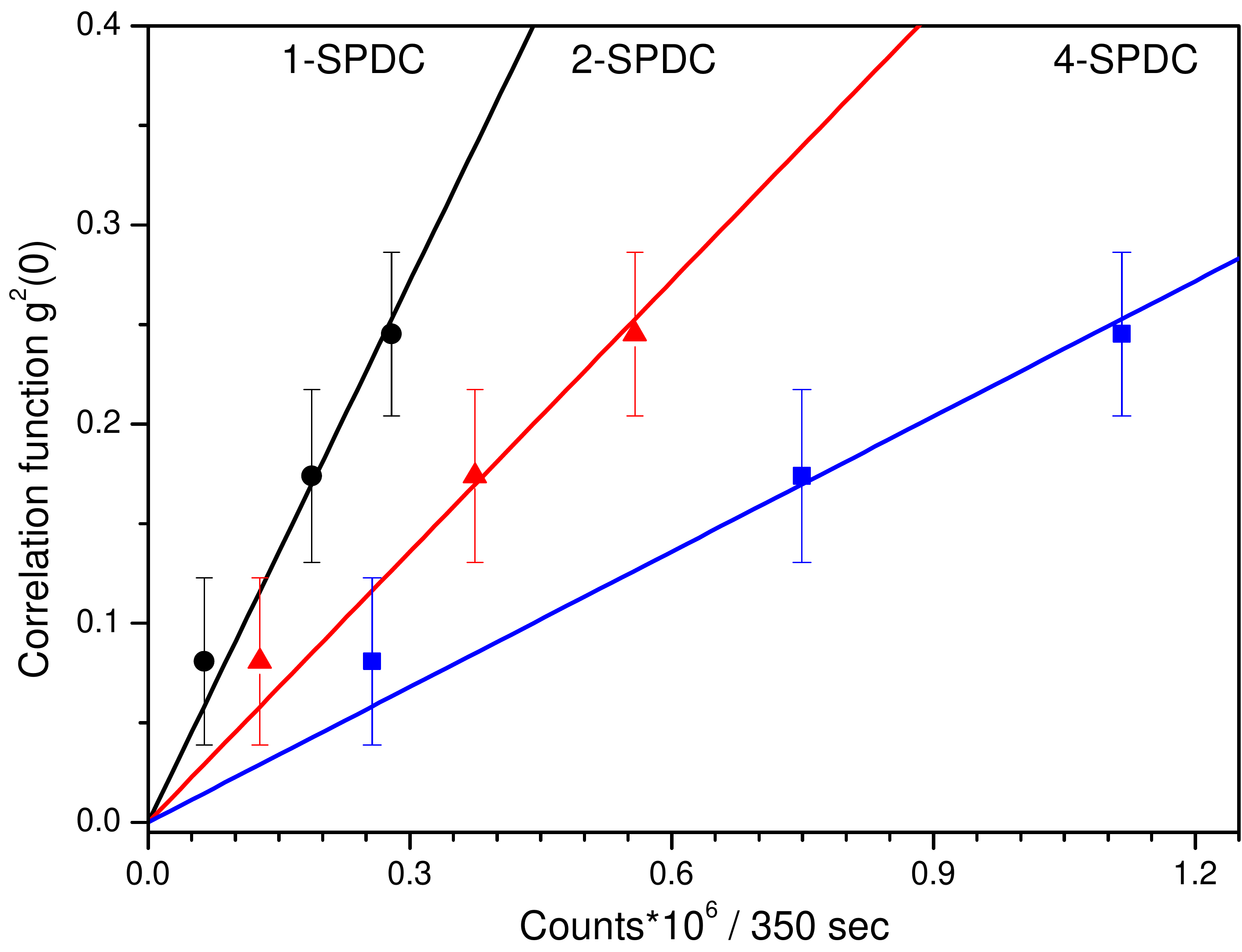}
\caption{\label{fig3}(Color online) Experimental results. The correlation function at zero
delay $g^{2}(0)$ is plotted versus the counts in 350 s for 1-SPDC (black circles), 2-SPDC (red triangles) and 4-SPDC (blue squares) multiplexed systems, where all of them contain the router circuitry. The $g^{2}(0)$ function saturates at the value 1 for large count rates, i.e. for large pump powers. The $g^{2}(0)$ data shown are in linear regime of low count rates and hence fitted with lines through the origin. Using multiplexed systems, one can increase the count rate while keeping constant the signal-to-noise ratio and hence the $g^{2}(0)$ function. In other words, increasing the number of SPDC sources enables one to increase the count rate with the same signal-to-noise ratio or (if one turns down the pump power) improves the quality of the single-photon output while maintaining the count rate of the output. Error bars indicate $\pm1$ standard deviation.}
\end{figure}

We quantified the quality of the multiplexed single photons with the second-order correlation function at zero time delay, $g^{2}(0)$. The smaller $g^{2}(0)$ is, the higher the quality of the single photons becomes. The correlation function can be measured in good approximation with a 50:50 BS and two more detectors
~\cite{Loudon1983}, as shown in Fig.\ 2c. It is defined as $g^{2}(0)=
N_{tTR}N_{t}/(N_{tT}N_{tR})$, where $N_{tTR}$, $N_{tT}$,
$N_{tR}$, and $N_{t}$  are the coincidence counts among trigger,
transmission, and reflection of the beam splitter, the coincidence
counts between trigger and transmission, the coincidence counts
between trigger and reflection, and the single counts of the trigger,
respectively. Figure \ref{fig3} shows $g^{2}(0)$ versus the
counts per 350 seconds with 1-SPDC, 2-SPDC and 4-SPDC
multiplexing systems.

We performed the experiments with various levels of pump power (100, 250, 500, and 1000 mW) to demonstrate the heralded single-photon quality and generation rate trade-off of the SPDC sources.
With a 1-SPDC source (black circles in Fig.\ \ref{fig3}), the counts increased as we increased the pump power. The quality of the output single photon decreased as $g^{2}(0)$ was increased because of higher order emissions. At a constant mean photon pair number $\bar{N}$ per source, with single-photon multiplexing, 2-SPDC (red triangles in Fig.\ \ref{fig3}) and 4-SPDC multiplexed (blue squares in Fig.\ \ref{fig3}) systems enhanced the counts by factors of 2 and 4, respectively, while $g^{2}(0)$ remained the same as the 1-SPDC. On the other hand, one can understand Fig.\ \ref{fig3} also from the following perspective: By accordingly decreasing $\bar{N}$ per source, one can employ multiplexed systems while keeping the counts constant and increase the quality of the output single photon via decreasing $g^{2}(0)$. In the regime of small $\bar{N}$, the counts scale proportionally to $m$ at fixed $g^{2}(0)$, and $g^{2}(0)$ scales inversely proportional to $m$ at fixed counts. Because the coupling efficiencies of the sources are different, the count rates and correlation functions are also different. The values of the counts and the correlation function shown in Fig.\ \ref{fig3} were averaged over four SPDC sources (see Appendix B).

\section*{Discussion}
The transmission throughput of the single-photon router is an important parameter for the viability of a multiplexed system, and next we provide an analysis of losses in our system. In our experiment, the output photon's transmission of the 4-SPDC multiplexing system is $T \approx 0.1$, where the main loss is due to single-mode-fiber to single-mode-fiber coupling in the router. To have an advantage of 4-SPDC multiplexing over a 1-SPDC without router circuitry would necessitate $T > 1/4$ or, equivalently, $T_0 > 1/2$ if the routers are all the same and each have a transmission of $T_0$. The gain due to multiplexing is $G \approx 4$ if the mean photon pair number is small (see Appendix A). Since the transmission of 0.1 of our router circuitry is a technical issue, our experiment is a proof-of-concept demonstration of single-photon multiplexing and demonstrates its advantage. Based on current technology and theoretical analysis of the fiber coupling efficiency, it may be feasible to improve the transmission per router to about 95\%. This number is obtained from experimentally achievable Fresnel losses on each optical surface and theoretically predicted fiber coupling efficiency~\cite{Ljunggren2005}. Then, the total gain of a 4-SPDC source with routers compared to a 1-SPDC source without router would be approximately $G\, T \approx 4\cdot 0.95^2 = 3.61$.

\begin{figure}
\includegraphics[width=0.47\textwidth]{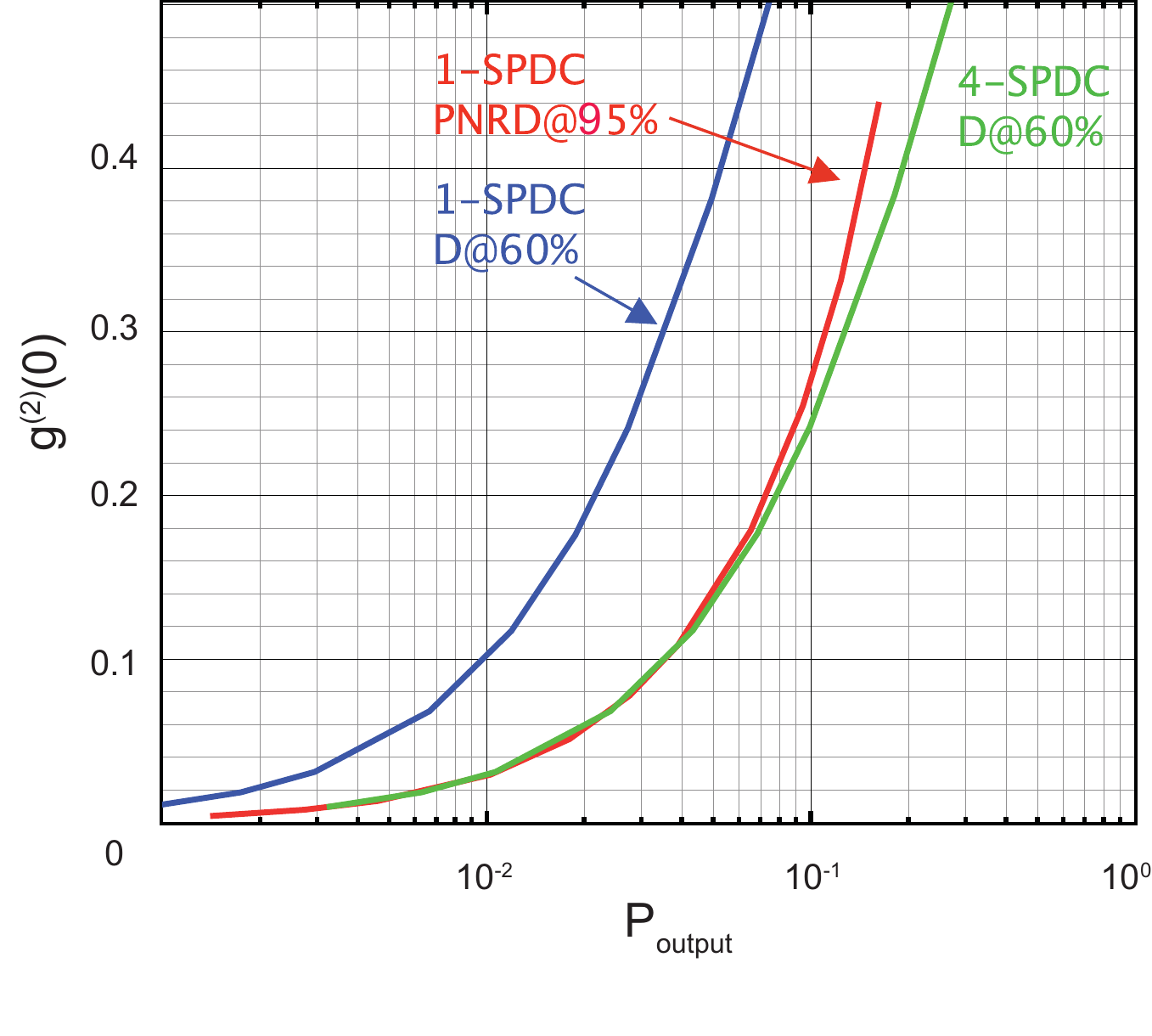}
\caption{\label{fig4}(Color online) Dependence of the photon correlation $g^{(2)}(0)$ vs the heralding output probability $\textrm{P}_{\textrm{output}}$, determined in a numerical simulation of SPDC based HSPSs with different types of detectors (see Appendix D and Ref.\ \cite{Jennewein2010}). The heralding output probability $\textrm{P}_\textrm{output}$ is defined as the ratio of the twofold coincidence counts (between trigger and output) and the laser repetition rate. The blue curve is for a 1-SPDC source with a standard bucket photon detector (D) with 60\% detection efficiency for the trigger photon. The red curve is assuming a photon-number-resolving detector (PNRD) with 95\% efficiency. Remarkably, as shown by the green curve, a 4-SPDC multiplexed system with standard detectors is able to outperform a 1-SPDC heralding source realized with a highly advanced PRND.}
\end{figure}

Practically, the repetition rate of multiplexing is only limited by the repetition rate of the pulsed laser (in our case 80 MHz, but femtosecond lasers with 1 GHz are commercially available~\cite{GIGA2010}), the jitter of the detector (500 ps in our case, but those with 50 ps are commercially available~\cite{ID2010}) and the switching time of the modulators (5.6 ns in our case, but less than 100 ps has been shown~\cite{Reed2010}). In addition, by employing state-of-the-art photonic structures, low insertion loss of the modulators can be achieved~\cite{Reed2010}. It is thus possible to realize a gigahertz on-demand single-photon source with this scheme.

Since in our experiment the photons are generated by the SPDC process pumped with a femtosecond-pulsed laser, which has been successfully used in multiphoton experiments in the past, a single photon source based on our scheme is suitable for high-quality multiphoton interference~\cite{Pan2008}, which is at the heart of the quantum repeater~\cite{Duan2001} and linear optical quantum computation~\cite{Kok2007}. It is very interesting to note that a 4-SPDC source such as demonstrated here based on standard bucket single-photon detectors, would potentially outperform a heralded photon source based on a highly efficient photon-number-resolving detector (PNRD) such as a superconducting transition-edge sensor (TES)~\cite{Lita2008} in specific multi-photon experiments~\cite{Jennewein2010} (see Fig.\ \ref{fig4} and Appendix D for details). This shows, that a 4-SPDC source would have important practical implications for advancing existing linear optical quantum computing experiments. It is conceivable that a scheme which combines the multiplexing technique with superconducting detectors as trigger detectors will be advantageous. Moreover, the scheme we have presented here is universal for any photon-pair source, including SPDC, spontaneous four-wave mixing, and so on.

For the future realization of a multiplexed system with many levels, stabilizing the interferometers will be challenging for bulk optics but less for integrated optics. Moreover, a 60\% transmission throughput of an integrated photonic quantum circuit has already been reported~\cite{Politi2008, Matthews2009}, which includes the fiber-coupling efficiency and is well above the loss threshold of $1/2$ mentioned above. Therefore, together with state-of-art micro-optics
technology, (for instance, integrated photonic quantum circuits on a
silicon chip~\cite{Politi2008, Matthews2009}, faster modulators
~\cite{Wooten2000}, and on-chip single-photon detectors~\cite{Kang2009}), it is possible to develop a compact, scalable and nearly on-demand single-photon source
for photon-based quantum science and technology following our scheme.

\begin{acknowledgements}
 We thank W. Naylor, S. Ramelow, C. Sch\"{a}ff and P. Walther for discussions. We acknowledge support from the European Commission, Project QAP (No.\ 015848), Q-ESSENCE (No.\ 248095), ERC advanced grant, SFB-FOQUS, JTF and the Doctoral Program CoQuS of the Austrian Science Foundation (FWF).
\end{acknowledgements}

\section*{Appendix}
\subsection{The model of an array of multiplexed heralded single photon sources}
In an array, $m$ \textit{sequentially} pumped HSPSs, based on SPDC, are coupled with $m-1$ photon routers. Since the array of SPDC sources is pumped sequentially, the source with the shortest path provides the first chance for getting a single photon, the next longer provides the next chance, and so on. The routers are operated in a way such that once they are activated by the trigger detector fired in a particular path, they block the outputs of all the other sources. Therefore, the source with the shortest path has "priority" in the array. The overall probability of obtaining one and only one output single photon out of the whole array is as follows:
\begin{equation}
Q_{1}=P_{1}+P_{0}P_{1}+...+P^{m-1}_{0}P_{1}=P_{1}\frac{1-P^{m}_{0}}{1-P_{0}},\label{Q1}
\end{equation}%
where $P_{0}$ is the probability of generating $0$ output photons from an individual SPDC source. Analogously, the probability of obtaining more than one output photon is:
\begin{equation}
Q_{>1}=P_{>1}\frac{1-P^{m}_{0}}{1-P_{0}}=(1-P_{0}-P_{1})\frac{1-P^{m}_{0}}{1-P_{0}}\label{Ql1}.
\end{equation}%
Therefore, the signal to noise ratio of obtaining one output single photon from this array equals that of a single SPDC source:
\begin{equation}
\frac{Q_{1}}{Q_{>1}}=\frac{P_{1}}{(1-P_{0}-P_{1})}=\frac{P_{1}}{P_{>1}}.\label{RQ}
\end{equation}%
The probability of obtaining one output single photon is enhanced by a gain factor of:
\begin{equation}
G=\frac{Q_{1}}{P_{1}}=\frac{1-P^{m}_{0}}{1-P_{0}}\approx\frac{1}{1-P_{0}},\label{G}
\end{equation}%
where the approximation at the end holds for $P^{m}_{0}\ll1$.

In any experimental setup, photon losses in the routers decrease the performance of the multiplexing scheme. We now analyze the influence of these losses quantitatively. We consider the case where the mean photon pair number $\bar{N}$ is much smaller than 1. Then, $P_0$ is close to 1 and thus the gain becomes $G \approx m$ [as long as $m(1-P_0)\ll1$]. Due to the cascaded setup, however, using $m$ SPDC sources means that the output photon has to travel through $n \equiv \log_2 m$ path routers (assuming $\log_2 m$ is an integer number) and hence has an increasing chance of being lost. The output photon's transmission through all $n$ path routers be denoted as $T$. In order to have an advantage over a single photon source, the gain $G$ has to overcompensate the loss such that the total gain, $G_{\textrm{tot}} \equiv G\, T$ is larger than 1. This leads to the generic condition $T > 1/m$. Then, $G_{\textrm{tot}}$ is increasing with $m$ and there is no upper bound. In case the transmission through all path routers is the same, and each single router transmission is denoted as $T_0$, the output photon's total transmission through the routers is $T = T_0^{\log_2 m}$. The total gain
\begin{equation}
G_{\textrm{tot}} \equiv G\, T \approx m\, T_0^{\log_2 m}
\end{equation}
has to be larger than 1, which leads to the condition that the transmission per path router must fulfill $T_0 > 1/2$. Therefore, multiplexing is advantageous over a single-photon source if and only if the loss of each individual path router is smaller than $1/2$ (under the assumption of small mean photon pair number). Losses in the routers do build up but are then overcompensated by the gain of the multiplexing.

In our hybrid experimental scheme, two different kinds of routers are needed, because we employ the polarization and the path degrees of freedom. In an $m$-SPDC arrangement utilizing polarization and path, there are only $\frac{m}{2}-1$ path routers (in our case, only one MZI because $m=4$), and it is sufficient to have one polarization router (in our scheme, EOM3) at the end of the array. Therefore, every output photon has to travel through $\log_2 \frac{m}{2}$ path routers (assuming $\log_2 \frac{m}{2}$ is an integer number) and one polarization router. When we denote the transmission of each path router by $T_{0}$ and the transmission of the polarization router by $T_{\textrm{pol}}$, the total transmission of the hybrid multiplexed system is $T_{\textrm{hybr}} = T_{0}^{\log_2 \frac{m}{2}}T_{\textrm{pol}}$. The total gain therefore becomes
\begin{equation}
G^{\textrm{hybr}}_{\textrm{tot}} \equiv G\, T_{\textrm{hybr}} \approx m\, T_{0}^{\log_2 \frac{m}{2}}T_{\textrm{pol}}.
\end{equation}

Demanding the total gain has to be larger than 1, leads to the condition $T_0 > \frac{1}{2 \sqrt[n]{2 T_{\textrm{pol}}}}$. In the case of a 4-SPDC multiplexed system with high-transmission polarization routing, $T_{\textrm{pol}} \approx 1$, and the requirement for the transmission of the path router is $T_0 > \frac{1}{4}$.

\subsection{The pulsed four-photon spontaneous parametric down conversion sources}
We used ultraviolet (UV) femtosecond pulses of an up-converted mode-locked Ti:sapphire laser to pump a 2 mm type-II BBO crystal at a wavelength of 404 nm with a pump power of 1400 mW, a pulse duration of 180 fs, and a repetition rate of 80 MHz. The UV pump power was adjusted by varying the up-conversion efficiency via moving the up-conversion crystal in or out of the focus of the lens for the fundamental pulse. The UV pulses successively passed through two BBO crystals to generate two polarization entangled photon pairs (photons 1 and 2, and photons 3 and 4) in spatial modes a and b, and c and d via type-II spontaneous parametric down-conversion respectively, as shown in Fig.\ 2(a). After proper filtering, maximally polarization entangled states of the form $\frac{1}{\sqrt{2}}
(\left\vert H\right\rangle _{\textrm{a(c)}}\left\vert V\right\rangle _{\textrm{b(d)}%
}-e^{i\theta}\left\vert V\right\rangle _{\textrm{a(c)}}\left\vert H\right\rangle _{\textrm{b(d)}%
}) $ were generated. $\theta$ is the phase difference between horizontal and vertical polarization due to birefringence in the crystal, which could be corrected by placing a half wave plate ($\lambda$/2) oriented at $45^{\circ}$ and 1-mm compensation BBO crystals in each photon's path~\cite{Kwiat1995}. By tilting the compensation BBO crystals, we set the phases to be 0 for both sources, and hence we obtained the two polarization-entangled states described in the text.

To exactly define the spatial and spectral properties of the emitted four photons, we coupled each photon into single-mode fibers and filter them with 3 nm (full width at half maximum, FWHM) bandpass interference filters (IF) for photons 1 and 4, and 1 nm IFs for photons 2 and 3. In the HOM interference experiment, we used 3 nm IFs for both photons. The photons were detected by single-photon counting modules (SPCMs), and the coincidences were registered with a coincidence logic. At the pump power of $1000$ mW, the average trigger rate is about $500$ kHz and the mean photon pair number generated per pulse is approximately $0.062$. Note that the coupling efficiency of the sources from BBO~1 is about $13$\%, and that from BBO~2 is about $7$\%. This difference eventually gives the difference in the correlation functions and the count rates of each source. For clarity, the 1-SPDC data shown in Fig.\ \ref{fig3} were obtained by averaging the $g^{2}(0)$ functions and the counts of the four individual 1-SPDC sources. The 2-SPDC (4-SPDC) data were obtained by averaging the $g^{2}(0)$ functions and summing the counts of two (four) 1-SPDC sources.

\subsection{The ultrafast photon router}
We used a MZI to realize the ultrafast photon router. The performance of this router strongly depends on the phase stability of the MZI. To achieve that, we built the interferometer in an enclosed box made by acoustic isolation materials in order to stabilize the phase passively. Additionally, an active phase stabilization system has also been implemented. This was accomplished by using an auxiliary beam from a power-stabilized He-Ne laser counter-propagating through the whole MZI with a little transversal displacement from the signal beam, which picked up the phase fluctuation of the interferometer. After passing through the interferometer, the intensity fluctuation of the He-Ne laser beam was measured with a silicon photon detector and the signal was fed into an analog proportional-integral-derivative (PID) regulator. A ring piezotransducer attached to one of the mirrors in the MZI was controlled by this PID regulator and compensated the phase fluctuation actively. The insertion loss of this router and fiber delays was about 90$\%$, which was mainly due to the single-mode fiber coupling, lossy narrow-band interference filters (in spatial modes b$''$ and c$''$, 1~nm FWHM, about 70\% peak transmission) and Fresnel loss on various optical elements. The most straight-forward step to improving the transmission of the router is to use high peak transmission and broadband interference filters. With a 95\% peak transmission and 3-nm FWHM interference filter, we have measured the total transmission throughput of the router, including the delay fibers and correlation function measurement setup, to be more than 25\%. Note that one can engineer the interactions between the pump and nonlinear crystals and match the group velocities of the pump and down-converted photons ~\cite{Mosley2008,Poh2009,Branczyk2010}. By employing this novel technique, one can generate pure single photons without using narrow-band filters, which potentially allows us to further enhance the transmission throughput to be more than 50\%.

\subsection{Comparing 4-SPDC to a 1-SPDC with different types of photon detectors}
In order to estimate the required performance of HSPS in practical applications, we implemented a numerical simulation of the quantum optics circuits. Thereby, every photon mode is represented in a Fock space from 0 to 4 excitations, and all unitary operations are connected Hamiltonians expressed within the full Hilbert space. Hence, all higher order photon terms are intrinsically taken into account, up to the natural cutoff from the Fock space. Through numerical simulation we studied the output probability for the HSPS versus the observed second-order correlation function for zero time delay, $g^{2}(0) = \frac{\langle (a^\dagger)^2 a^2 \rangle}{\langle a^\dagger a \rangle^2}$, for various pumping strengths $\epsilon^2$ of the SPDC operator, $H_{\mathrm{SPDC}}=\epsilon(a^\dagger b^\dagger + b a),$ where $a$ and $b$ are the two photon modes. These results are shown in Fig.\ \ref{fig4}.
Interestingly, the 4-SPDC source based on 60$\%$ efficiency bucket detectors (for detecting trigger photons) and optical transmission throughput of 95\% per router, which should be reasonable to reach with an optimized optical system, can outperform a HSPS built from 1-SPDC heralded with the currently best possible detector technology, such as a PNRD system (for detecting trigger photons) with total detector efficiency of 95$\%$. Note that we assume the coupling efficiency from the source to the detectors to be 70\% for both cases, because that is independent of the detectors.


\begin{thebibliography}{10}
\expandafter\ifx\csname url\endcsname\relax
  \def\url#1{\texttt{#1}}\fi
\expandafter\ifx\csname urlprefix\endcsname\relax\def\urlprefix{URL }\fi
\providecommand{\bibinfo}[2]{#2}
\providecommand{\eprint}[2][]{\url{#2}}

\bibitem{Gisin2002}
\bibinfo{author}{Gisin, N.}, \bibinfo{author}{Ribordy, G.},
  \bibinfo{author}{Tittel, W.} \& \bibinfo{author}{Zbinden, H.}
\newblock \emph{\bibinfo{journal}{Rev. Mod. Phys.}}
  \textbf{\bibinfo{volume}{74}}, \bibinfo{pages}{145--195}
  (\bibinfo{year}{2002}).

\bibitem{Scarani2009}
\bibinfo{author}{Scarani V.} \emph{et~al.}
\newblock \emph{\bibinfo{journal}{Rev. Mod. Phys.}}
  \textbf{\bibinfo{volume}{81}}, \bibinfo{pages}{1301--1350}
  (\bibinfo{year}{2009}).

\bibitem{Bouwmeester1997}
\bibinfo{author}{Bouwmeester D.} \emph{et~al.}
\newblock \emph{\bibinfo{journal}{Nature}} \textbf{\bibinfo{volume}{390}},
  \bibinfo{pages}{575--579} (\bibinfo{year}{1997}).


\bibitem{Pan1998}
\bibinfo{author}{Pan, J.-W.}, \bibinfo{author}{Bouwmeester, D.},
  \bibinfo{author}{Weinfurter, H.} \& \bibinfo{author}{Zeilinger, A.}
\newblock \emph{\bibinfo{journal}{Phys. Rev. Lett.}}
  \textbf{\bibinfo{volume}{80}}, \bibinfo{pages}{3891--3894}
  (\bibinfo{year}{1998}).

\bibitem{Jennewein2002}
\bibinfo{author}{Jennewein, T.}, \bibinfo{author}{Weihs, G.},
  \bibinfo{author}{Pan, J.-W.} \& \bibinfo{author}{Zeilinger, A.}
\newblock \emph{\bibinfo{journal}{Phys. Rev. Lett.}}
  \textbf{\bibinfo{volume}{88}}, \bibinfo{pages}{017903}
  (\bibinfo{year}{2001}).


\bibitem{Kok2007}
\bibinfo{author}{Kok P.} \emph{et~al.}
\newblock \emph{\bibinfo{journal}{Rev. Mod. Phys.}}
  \textbf{\bibinfo{volume}{79}}, \bibinfo{pages}{135} (\bibinfo{year}{2007}).


\bibitem{Lounis2000}
\bibinfo{author}{Lounis, B.} \& \bibinfo{author}{Moerner, W.~E.}
\newblock \emph{\bibinfo{journal}{Nature}} \textbf{\bibinfo{volume}{407}},
  \bibinfo{pages}{491--493} (\bibinfo{year}{2000}).

\bibitem{Lettow2010}
\bibinfo{author}{Lettow R.} \emph{et~al.}
\newblock \emph{\bibinfo{journal}{Phys. Rev. Lett.}}
  \textbf{\bibinfo{volume}{104}}, \bibinfo{pages}{123605}
  (\bibinfo{year}{2010}).

\bibitem{Chou2004}
\bibinfo{author}{Chou, C.~W.}, \bibinfo{author}{Polyakov, S.~V.},
  \bibinfo{author}{Kuzmich, A.} \& \bibinfo{author}{Kimble, H.~J.}
\newblock \emph{\bibinfo{journal}{Phys. Rev. Lett.}}
  \textbf{\bibinfo{volume}{92}}, \bibinfo{pages}{213601}
  (\bibinfo{year}{2004}).

\bibitem{Kurtsiefer2000}
\bibinfo{author}{Kurtsiefer, C.}, \bibinfo{author}{Mayer, S.},
  \bibinfo{author}{Zarda, P.} \& \bibinfo{author}{Weinfurter, H.}
\newblock \emph{\bibinfo{journal}{Phys. Rev. Lett.}}
  \textbf{\bibinfo{volume}{85}}, \bibinfo{pages}{290--293}
  (\bibinfo{year}{2000}).

\bibitem{Michler2000}
\bibinfo{author}{Michler M.} \emph{et~al.}
\newblock \emph{\bibinfo{journal}{Science}}
  \textbf{\bibinfo{volume}{290}}, \bibinfo{pages}{2282--2285}
  (\bibinfo{year}{2000}).

\bibitem{Santori2002}
\bibinfo{author}{Santori, C.}, \bibinfo{author}{Fattal, D.},
  \bibinfo{author}{Vuckovic, J.}, \bibinfo{author}{Solomon, G.~S.} \&
  \bibinfo{author}{Yamamoto, Y.}
\newblock \emph{\bibinfo{journal}{Nature}} \textbf{\bibinfo{volume}{419}},
  \bibinfo{pages}{594--597} (\bibinfo{year}{2002}).

\bibitem{Sanaka2009}
\bibinfo{author}{Sanaka, K.}, \bibinfo{author}{Pawlis, A.},
  \bibinfo{author}{Ladd, T.~D.}, \bibinfo{author}{Lischka, K.} \&
  \bibinfo{author}{Yamamoto, Y.}
\newblock \emph{\bibinfo{journal}{Phys. Rev. Lett.}}
  \textbf{\bibinfo{volume}{103}}, \bibinfo{pages}{053601}
  (\bibinfo{year}{2009}).

\bibitem{Batalov2008}
\bibinfo{author}{Batalov A.} \emph{et~al.}
\newblock \emph{\bibinfo{journal}{Phys. Rev. Lett.}}
  \textbf{\bibinfo{volume}{100}}, \bibinfo{pages}{077401}
  (\bibinfo{year}{2008}).

\bibitem{Patel2010}
\bibinfo{author}{Patel R.~B.} \emph{et~al.}
\newblock \emph{\bibinfo{journal}{Nat. Photon}}  \textbf{\bibinfo{volume}{4}},
  \bibinfo{pages}{632--635} (\bibinfo{year}{2010}).

\bibitem{Shapiro2007}
\bibinfo{author}{Shapiro, J.~H.} \& \bibinfo{author}{Wong, F.~N.}
\newblock \emph{\bibinfo{journal}{Opt. Lett.}} \textbf{\bibinfo{volume}{32}},
  \bibinfo{pages}{2698--2700} (\bibinfo{year}{2007}).

\bibitem{Migdall2002}
\bibinfo{author}{Migdall, A.~L.}, \bibinfo{author}{Branning, D.} \&
  \bibinfo{author}{Castelletto, S.}
\newblock \emph{\bibinfo{journal}{Phys. Rev. A}} \textbf{\bibinfo{volume}{66}},
  \bibinfo{pages}{053805} (\bibinfo{year}{2002}).

\bibitem{Pittman2002}
\bibinfo{author}{Pittman, T.~B.}, \bibinfo{author}{Jacobs, B.~C.} \&
  \bibinfo{author}{Franson, J.~D.}
\newblock \emph{\bibinfo{journal}{Phys. Rev. A}} \textbf{\bibinfo{volume}{66}},
  \bibinfo{pages}{042303} (\bibinfo{year}{2002}).

\bibitem{Jeffrey2004}
\bibinfo{author}{Jeffrey, E.}, \bibinfo{author}{Peters, N.~A.} \&
  \bibinfo{author}{Kwiat, P.~G.}
\newblock \emph{\bibinfo{journal}{New J. Phys.}} \textbf{\bibinfo{volume}{6}},
  \bibinfo{pages}{100} (\bibinfo{year}{2004}).

\bibitem{McCusker2009}
\bibinfo{author}{McCusker, K.~T.}, \& \bibinfo{author}{Kwiat, P.~G.}
\newblock \emph{\bibinfo{journal}{Phys. Rev. Lett.}}
  \textbf{\bibinfo{volume}{103}}, \bibinfo{pages}{163602}
  (\bibinfo{year}{2009}).

\bibitem{Ladd2010}
\bibinfo{author}{Ladd T.~D.} \emph{et~al.}
\newblock \emph{\bibinfo{journal}{Nature}} \textbf{\bibinfo{volume}{464}},
  \bibinfo{pages}{45--53} (\bibinfo{year}{2010}).


\bibitem{Kok2010}
\bibinfo{author}{Kok, P.}
\newblock \emph{\bibinfo{journal}{Nat. Photon}}  \textbf{\bibinfo{volume}{4}},
\bibinfo{pages}{504--505} (\bibinfo{year}{2010}).


\bibitem{Kwiat1995}
\bibinfo{author}{Kwiat P.~G.} \emph{et~al.}
\newblock \emph{\bibinfo{journal}{Phys. Rev. Lett.}}
  \textbf{\bibinfo{volume}{75}}, \bibinfo{pages}{4337--4341}
  (\bibinfo{year}{1995}).

\bibitem{Treiber2009}
\bibinfo{author}{Treiber A.} \emph{et~al.}
\newblock \emph{\bibinfo{journal}{New J. Phys.}} \textbf{\bibinfo{volume}{11}},
  \bibinfo{pages}{045013} (\bibinfo{year}{2009}).

\bibitem{Bennett1984}
\bibinfo{author}{Bennett, C. H.} \& \bibinfo{author}{Brassard, G.} \newblock in \emph{\bibinfo{journal}{Proceedings of IEEE
International Conference on Computers, Systems, and
Signal Processing, Bangalore, India}},
  \bibinfo{pages}{175} (\bibinfo{year}{1984}).

\bibitem{Reed2010}
\bibinfo{author}{Reed, G.~T.}, \bibinfo{author}{Mashanovich, G.},
  \bibinfo{author}{Gardes, F.~Y.} \& \bibinfo{author}{Thomson, D.~J.}
\newblock \emph{\bibinfo{journal}{Nat. Photon}} \textbf{\bibinfo{volume}{4}},
  \bibinfo{pages}{518--526} (\bibinfo{year}{2010}).

\bibitem{Scheidl2008}
\bibinfo{author}{Scheidl T.} \emph{et~al.}
\newblock \emph{\bibinfo{journal}{PNAS}} \textbf{\bibinfo{volume}{107}}, \bibinfo{pages}{19708-19713} (\bibinfo{year}{2010}).

\bibitem{Hong1987}
\bibinfo{author}{Hong, C.~K.}, \bibinfo{author}{Ou, Z.~Y.} \&
  \bibinfo{author}{Mandel, L.}
\newblock \emph{\bibinfo{journal}{Phys. Rev. Lett.}}
  \textbf{\bibinfo{volume}{59}}, \bibinfo{pages}{2044--2046}
  (\bibinfo{year}{1987}).

\bibitem{Kaltenbaek2006}
\bibinfo{author}{Kaltenbaek, R.}, \bibinfo{author}{Blauensteiner, B.},
  \bibinfo{author}{\.{Z}ukowski, M.}, \bibinfo{author}{Aspelmeyer, M.} \&
  \bibinfo{author}{Zeilinger, A.}
\newblock \emph{\bibinfo{journal}{Phys. Rev. Lett.}}
  \textbf{\bibinfo{volume}{96}}, \bibinfo{pages}{240502}
  (\bibinfo{year}{2006}).

\bibitem{Halder2007}
\bibinfo{author}{Halder M.} \emph{et~al.}
\newblock \emph{\bibinfo{journal}{Nat. Phys.}} \textbf{\bibinfo{volume}{3}},
  \bibinfo{pages}{692--695} (\bibinfo{year}{2007}).

\bibitem{Pan2008}
\bibinfo{author}{Pan, J.-W.}, \bibinfo{author}{Chen, Z.-B.},
  \bibinfo{author}{Zukowski, M.}, \bibinfo{author}{Weinfurter, H.} \&
  \bibinfo{author}{Zeilinger, A.}
\newblock \emph{\bibinfo{journal}{arXiv}} \bibinfo{pages}{0805.2853}
  (\bibinfo{year}{2008}).

\bibitem{Langford2005}
\bibinfo{author}{Langford, N.~K.} \emph{et~al.}
\newblock \emph{\bibinfo{journal}{Phys. Rev. Lett.}}
  \textbf{\bibinfo{volume}{95}}, \bibinfo{pages}{210504}
  (\bibinfo{year}{2005}).

\bibitem{Kiesel2005}
\bibinfo{author}{Kiesel, N.}, \bibinfo{author}{Schmid, C.},
  \bibinfo{author}{Weber, U.}, \bibinfo{author}{Ursin, R.} \&
  \bibinfo{author}{Weinfurter, H.}
\newblock \emph{\bibinfo{journal}{Phys. Rev. Lett.}}
  \textbf{\bibinfo{volume}{95}}, \bibinfo{pages}{210505}
  (\bibinfo{year}{2005}).

\bibitem{Okamoto2005}
\bibinfo{author}{Okamoto, R.}, \bibinfo{author}{Hofmann, H.~F.},
  \bibinfo{author}{Takeuchi, S.} \& \bibinfo{author}{Sasaki, K.}
\newblock \emph{\bibinfo{journal}{Phys. Rev. Lett.}}
  \textbf{\bibinfo{volume}{95}}, \bibinfo{pages}{210506}
  (\bibinfo{year}{2005}).

\bibitem{Loudon1983}
\bibinfo{author}{Loudon, R.}
\newblock \emph{\bibinfo{title}{The quantum theory of light}}
  (\bibinfo{publisher}{Oxford Univeristy Press}, \bibinfo{year}{1983}).

\bibitem{Ljunggren2005}
\bibinfo{author}{Ljunggren, D.} \& \bibinfo{author}{Tengner, M.}
\newblock \emph{\bibinfo{journal}{Phys. Rev. A}} \textbf{\bibinfo{volume}{72}},
  \bibinfo{pages}{062301} (\bibinfo{year}{2005}).

\bibitem{GIGA2010}
http://www.gigaoptics.de

\bibitem{ID2010}
http://www.idquantique.com/.

\bibitem{Duan2001}
\bibinfo{author}{Duan, L.~M.}, \bibinfo{author}{Lukin, M.~D.},
  \bibinfo{author}{Cirac, J.~I.} \& \bibinfo{author}{Zoller, P.}
\newblock \emph{\bibinfo{journal}{Nature}} \textbf{\bibinfo{volume}{414}},
  \bibinfo{pages}{413--418} (\bibinfo{year}{2001}).

\bibitem{Lita2008}
\bibinfo{author}{Lita, A.~E.}, \bibinfo{author}{Miller, A.~J.} \&
  \bibinfo{author}{Nam, S.~W.}
\newblock \emph{\bibinfo{journal}{Opt. Express}} \textbf{\bibinfo{volume}{16}},
  \bibinfo{pages}{3032--3040} (\bibinfo{year}{2008}).

\bibitem{Jennewein2010}
\bibinfo{author}{Jennewein, T.}, \bibinfo{author}{Barbieri, M.} \&
  \bibinfo{author}{White, A.}
\newblock \emph{\bibinfo{journal}{arXiv}} \bibinfo{pages}{1012.1868}
  (\bibinfo{year}{2010}).

\bibitem{Politi2008}
\bibinfo{author}{Politi, A.}, \bibinfo{author}{Cryan, M.~J.},
  \bibinfo{author}{Rarity, J.~G.}, \bibinfo{author}{Yu, S.} \&
  \bibinfo{author}{O'Brien, J.~L.}
\newblock \emph{\bibinfo{journal}{Science}} \textbf{\bibinfo{volume}{320}},
  \bibinfo{pages}{646--649} (\bibinfo{year}{2008}).

\bibitem{Matthews2009}
\bibinfo{author}{Matthews, J. C.~F.}, \bibinfo{author}{Politi, A.},
  \bibinfo{author}{StefanovAndre} \& \bibinfo{author}{O'Brien, J.~L.}
\newblock \emph{\bibinfo{journal}{Nat. Photon}} \textbf{\bibinfo{volume}{3}},
  \bibinfo{pages}{346--350} (\bibinfo{year}{2009}).

\bibitem{Wooten2000}
\bibinfo{author}{Wooten, E.} \emph{et~al.}
\newblock \emph{\bibinfo{journal}{IEEE J. Sel. Top. Quant. Electron.}} \textbf{\bibinfo{volume}{6}}, \bibinfo{pages}{69 --82}
  (\bibinfo{year}{2000}).

\bibitem{Kang2009}
\bibinfo{author}{Kang, Y.} \emph{et~al.}
\newblock \emph{\bibinfo{journal}{Nat Photon}} \textbf{\bibinfo{volume}{3}},
  \bibinfo{pages}{59--63} (\bibinfo{year}{2009}).

\bibitem{Mosley2008}
\bibinfo{author}{Mosley, P. J.} \emph{et~al.}
\newblock \emph{\bibinfo{journal}{Phys. Rev. Lett}} \textbf{\bibinfo{volume}{100}},
  \bibinfo{pages}{133601} (\bibinfo{year}{2008}).

\bibitem{Poh2009}
\bibinfo{author}{Poh, H. S.}, \bibinfo{author}{Lim, J.},
  \bibinfo{author}{Marcikic, I.}, \bibinfo{author}{Lamas-Linares, A.} \&
  \bibinfo{author}{Kurtsiefer, C.}
\newblock \emph{\bibinfo{journal}{Phys. Rev. A}} \textbf{\bibinfo{volume}{80}},
  \bibinfo{pages}{043815} (\bibinfo{year}{2009}).

\bibitem{Branczyk2010}
\bibinfo{author}{Bra{\'n}czyk, A. M.}, \bibinfo{author}{Fedrizzi, A.},
\bibinfo{author}{Ralph, T. C.}, \& \bibinfo{author}{White, A. G.},
\newblock \emph{\bibinfo{journal}{arxiv}} \bibinfo{pages}{1005.3086} (\bibinfo{year}{2010}).

\end{thebibliography}
\end{document}